# London superconductivity approach in a time-dependent background

V. Aguiar*, J. P. G. Nascimento, I. Guedes and R. N. Costa Filho

Departamento de Física, Universidade Federal do Ceará, Campus do PICI, Caixa Postal 6030, 60455-760, Fortaleza, CE, Brazil.

**Abstract**

The main goal of this paper is to obtain the exact quantum solutions for charge space in a superconductor with time-dependent parameters using the London approach. We introduce a new quantization scheme for the charge inside a superconductor based on the Lewis and Riesenfeld invariant operator method. From the wave-functions obtained, we calculated the time-dependent uncertainties and the mean energy of the system. Information measures were also obtained, such as Shannon entropy and complexity. The later is always time-independent and also does not depend on conductivity. The others quantities are written in terms of a time-dependent function, $\rho(t)$, c-number quantity satisfying a nonlinear differential equation.

*Corresponding author. Tel.: +55 85998436604.

E-mail address: vanderley.junior.ufc@gmail.com (V. Aguiar).

# 1. Introduction

The miniaturization of electronic devices and integrated circuits to nanoscale has attracted a great deal of attention in the literature, owing to the advance of mesoscopic physics and nanotechnology [1-7]. Consequently, these devices (or circuits) are so small that the inelastic coherence of the charge carrier approaches the Fermi wavelength, so the application of classical mechanics fails and quantum effects have to be considered. The most simple mesoscopic circuit is the LC (inductance-capacitance) one, which is a nondissipative quantum system.

The first quantization scheme of an LC circuit was presented by W. H. Louisell [8] in the 1970s as a firt step to study the problem of quantizing an electromagnetic field in a cavity. He analyzed the quantum effects of a nondissipative quantum LC circuit with a source and expressed its fluctuations in the vacuum state. Since then, the quantum effects of these mesoscopics circuits were extensively studied in both theorethical and experimental point of view. For instance, Zhang et al. [9] presented a quantization scheme for an RLC (R, L, and C are constants) circuit with a source and studied the fluctuations of the charge and magnetic flux of the circuit in several quantum states. In a series of papers, Pedrosa and coworkers [10-13] employed the Lewis–Riesenfeld method [14] to obtain the solution of the Schrödinger equation for time-dependent RLC circuits and constructed their coherent and squeezed states. Aguiar, Guedes and Pedrosa [15] calculated the time-dependent Shannon, Tsallis and Rényi entropies of time-dependent RLC circuit.

Recently, mesoscopic LC circuits coupled with Josephson junctions have attracted considerable attention in the literature because of their possible applicability in quantum computing. A Josephson junction is composed by two superconductors weakly connected by a thin layer of insulating material [16]. For example, Meng, Wang and Liang [16] obtained the classical Hamiltonian for the mesoscopic LC circuit including two coupled Josephson junctions using Louisell quantization. Wang, Liang and Fan [17] showed that when a single Josephson junction is inserted in a mesoscopic LC circuit, the second Josephson equation is modified due to the variation of the magnetic flux in the inductance. Other interesting results about mesoscopic LC circuits including Josephson junctions using the Louisell quantization scheme in the charge continuum and discreteness condition can be found in Refs. [18-21].

The Louisell quantization was constructed utilizing the classical analogy between a nondissipative LC circuit and the simple harmonic oscillator, in which the classical variables $x \to q$ (charge) and $p \to I$ (current), thus the Hamiltonian of the circuit can be express as

$$H_{LC} = \frac{LI^2}{2} + \frac{q^2}{2C}, \quad (1)$$

where $L$ and $C$ is the inductance and capacitance of the circuit, respectively. One can show that the classical "equation of motion" is given by

$$\ddot{q} + \omega_0^2 q = 0, \quad (2)$$

where $\omega_0 = \sqrt{\frac{1}{LC}}$ is the circuit resonant frequency. To obtain Eq. (2) using Hamiltonian mechanics, the phase-space coordinates must be charge $q$ and magnetic flux $\Phi$, contrary to what was initially proposed by Louisell. In this way, the quantization of this system is based in two non-commuting Hermitian operators, the charge operator $q$ and magnetic flux operator $\Phi = -i\hbar \frac{\partial}{\partial q}$, which satisfy the commutation relation $[q, \Phi] = i\hbar$.

Although the importance of the quantum LC circuit in theoretical physics point of view, in more practical circuits, a mesoscopic resistance must be taken into account. Thus, in an RLC mesoscopic circuit, the classical equation of motion for the charge is given by

$$\ddot{q} + \frac{R}{L}\dot{q} + \omega_0^2 q = 0, \quad (3)$$

which is identical to the equation of motion of a damped harmonic oscillator. This system was extensively studied in the last decade [9-13, 15, 22-24]. However, an interesting parallel went unnoticed.

This analogy lies in the form of Eq. (3) for the charge. To illustrate it, we must go back to the early 1930s, when the two brothers Heinz and Fritz London [25] introduced in the literature a phenomenological approach to explain some fundamental properties of superconductors, in particular, the expulsion of a magnetic field from a superconductor, the so-called Meissner-Ochsenfeld effect [26].

Although the London approach is based on very simple arguments, it explains satisfactorily certain properties of the superconductors. Consider the London equations

$$\frac{\partial \boldsymbol{J}_s}{\partial t} = \frac{n_s e^2}{m_e} \boldsymbol{E}, \qquad (4)$$

$$\nabla \times \boldsymbol{J}_s = -\frac{n_s e^2}{m_e c} \boldsymbol{B}, \qquad (5)$$

where $n_s$ is the density of electrons in a superconducing state and $\boldsymbol{J}_s$ is the supercurrent.

Next, in addition to superfluid electrons there are normal electrons, giving rise to a normal current [27]

$$\boldsymbol{J}_n = \sigma(t) \boldsymbol{E}, \qquad (6)$$

where $\sigma(t)$ is a time-dependent conductivity. So the total current reads

$$\boldsymbol{J} = \boldsymbol{J}_s + \boldsymbol{J}_n. \qquad (7)$$

Using Eqs. (4), (6) and (7) toghether with the continuity and Maxwell equations, we obtain the following equation for a charge within a superconductor as

$$\ddot{q} + \frac{\sigma(t)}{\varepsilon_0} \dot{q} + \left(\frac{c^2}{\lambda_L^2} + \frac{\dot{\sigma}}{\varepsilon_0}\right) q = 0, \qquad (8)$$

where $\lambda_L = \frac{m_e}{\mu_0 n_s e^2}$ is the London penetration depth.

Equation (8) is identical to the Eq. (3), but with time-dependent parameters and it can be obtained from the explicitly time-dependent Hamiltonian

$$H(q, \Phi, t) = e^{-\int \frac{\sigma(t)}{\varepsilon_0} dt} \frac{\Phi^2}{2} + \left(\frac{c^2}{\lambda_L^2} + \frac{\dot{\sigma}}{\varepsilon_0}\right) \frac{q^2}{2} e^{\int \frac{\sigma(t)}{\varepsilon_0} dt}, \qquad (9)$$

where the charge $q$ and the magnetic flux $\Phi$ are the canonical variables.

The quantization of this problem is written in terms of the conjugate operators charge $q$ and magnetic flux $\Phi = -i\hbar \frac{\partial}{\partial q}$. Thus, in terms of $q$ and $\Phi$ the time-dependent Schrödinger equation reads

$$i\hbar \frac{\partial \psi(q,t)}{\partial t} = \left[ -e^{-\int \frac{\sigma(t)}{\varepsilon_0} dt} \frac{\hbar^2}{2} \frac{\partial^2}{\partial q^2} + \frac{\omega^2(t)}{2} q^2 e^{\int \frac{\sigma(t)}{\varepsilon_0} dt} \right] \psi(q,t), \qquad (10)$$

where $\omega^2(t) = \left(\frac{c^2}{\lambda_L^2} + \frac{\dot{\sigma}}{\varepsilon_0}\right)$.

The quantum solutions of this problem, $\psi(q,t)$, can be exactly obtained by applying the Lewis and Riesenfeld invariant method [14]. Therefore the main goal of this paper is to obtain the exact quantum solutions of Eq. (10) for charge in a superconductor using the London approach and calculate the uncertainties and information measures for the systems. In Section 2 we briefly explain the fundamental definitions needed for the calculations. In Section 3, we present the calculations and results. The concluding remarks are presented in Section 4.

## 2. Theory

It is well known that an invariant operator $I(t)$ satisfies the equation

$$\frac{dI}{dt} = \frac{\partial I}{\partial t} + \frac{1}{i\hbar}[I, H] = 0, \qquad (11)$$

and for Eq. (10) it is given by [14]

$$I(t) = \frac{1}{2}\left[\left(\frac{q}{\rho}\right)^2 + (\rho\Phi - L\dot{\rho}q)^2\right], \qquad (12)$$

where $q(t)$ satisfies Eq. (5), $L(t) = e^{\int \frac{\sigma(t)}{\varepsilon_0} dt}$ and $\rho(t)$ satisfies the generalized Milne-Pinney [28, 29] equation

$$\ddot{\rho} + \frac{\dot{L}}{L}\dot{\rho} + \omega^2(t)\rho = \frac{1}{L^2(t)\rho^3}. \qquad (13)$$

The invariant operator $I(t)$ will be hermitian only for real solutions of Eq. (13). Its eigenfunctions, $\phi_n(q,t)$, are assumed to form a complete orthonormal set with time-independent discrete eigenvalues, $\lambda_n$. Thus

$$I\phi_n(q,t) = \lambda_n \phi_n(q,t), \qquad (14)$$

with $\langle \phi_n | \phi_{n'} \rangle = \delta_{nn'}$.

The solutions $\psi_n(q,t)$ of Eq. (10) are related to the functions $\phi_n(q,t)$ by

$$\psi_n(q,t) = e^{i\theta_n(t)} \phi_n(q,t), \qquad (15)$$

where the phase functions $\theta_n(t)$ satisfy the equation

$$\theta_n(t) = -\left(n + \frac{1}{2}\right) \int_{t_0}^{t} \frac{1}{L(t')\rho^2} dt'. \qquad (16)$$

Next, consider the unitary transformation

$$\phi'_n(q,t) = U\phi_n(q,t), \tag{17}$$

where

$$U = \exp\left\{-\left[i\frac{L(t)\dot{\rho}}{2\hbar\rho}\right]q^2\right\}. \tag{18}$$

Under this transformation and defining $\xi = \frac{q}{\rho}$, Eq. (14) now reads

$$I'\varphi_n(\xi) = \left[-\left(\frac{\hbar^2}{2}\right)\frac{\partial^2}{\partial\xi^2} + \left(\frac{\xi^2}{2}\right)\right]\varphi_n(\xi) = \lambda_n\varphi_n(\xi), \tag{19}$$

where $I' = UIU^\dagger$, $\frac{\varphi_n(\xi)}{\rho^{1/2}} = \phi'_n(q,t)$ and $\lambda_n = \left(n+\frac{1}{2}\right)\hbar$. The factor $\rho^{1/2}$ warrants the normalization condition

$$\int {\phi'_n}^*(q,t)\phi'_n(q,t)dq = \int \varphi_n^*(\xi)\varphi_n(\xi)d\xi = 1. \tag{20}$$

Since the solution of Eq. (19) corresponds to that of the time-independent harmonic oscillator, the solutions of Eq. (10) read

$$\psi_n(q,t) = e^{i\theta_n(t)}\left[\frac{1}{\pi^{1/2}\hbar^{1/2}n!\,2^n\rho}\right]^{1/2}\exp\left[\frac{iL(t)}{2\hbar}\left(\frac{\dot{\rho}}{\rho} + \frac{i}{L(t)\rho^2}\right)q^2\right]$$

$$\times H_n\left[\left(\frac{1}{\hbar}\right)^{1/2}\left(\frac{q}{\rho}\right)\right], \tag{21}$$

where we used Eqs. (15), (16), (17), (18) and (20), and $H_n$ is the usual Hermite polynomial of order $n$.

From Eq. (21) the following average values and uncertainty relation can be found

$$\langle q \rangle = \langle \Phi \rangle = 0, \qquad (22)$$

$$\langle q^2 \rangle = \hbar \rho^2 \left(n + \frac{1}{2}\right), \qquad (23)$$

$$\langle \Phi^2 \rangle = \frac{\hbar}{\rho^2}(1 + L^2(t)\rho^2 \dot{\rho}^2)\left(n + \frac{1}{2}\right), \qquad (24)$$

$$\Delta q \Delta \Phi = \hbar(1 + L(t)^2 \rho^2 \dot{\rho}^2)^{\frac{1}{2}}\left(n + \frac{1}{2}\right). \qquad (25)$$

The Eqs. (22)-(25) give us the time-dependent behavior of the quantization of the charge ($q$) and the magnetic flux ($\Phi$) inside a superconductor due to the normal electrons that give rise to the normal current Eq. (6).

Although the Heisenberg uncertainty relations play an important role in quantum mechanics, new forms of uncertainty relations that may lead to limits beyond those given by the Heisenberg uncertainty relation have been taken into account. For instance, we mention the entropic uncertainty relations obtained from the Shannon entropy [30-33] and the generalized and extendend uncertainty principle [34, 35].

The Shannon entropy ($S_q$) [36] for the observable $q$ are given by

$$S_q = -\int \mathcal{P}(q,t) \ln[\mathcal{P}(q,t)] \, dq, \qquad (26)$$

and provide useful uncertainty measures with respect to the probability density $\mathcal{P}(q,t) = |\psi_n(q,t)|^2$.

Over the past decades, the Shannon entropy have also been used to derive different kinds of statistical measures such as the statistical complexity [37, 38]. The statistical complexity is defined as

$$C_q = H_q \cdot D_q, \tag{27}$$

where $H_q = e^{S_q}$ and $D_q$ is the so called disequilibrium

$$D_q = \int [\mathcal{P}(q,t)]^2 dq, \tag{28}$$

which measures how far is the distribution from its equiprobability. These quantities have been used to study several physical phenomena. In 2014, Sañudo et al [39] obtained the statistical complexity for electrons scattering across a potential barrier in a monolayer graphene. They observed that the statistical complexity take its minimum value in the situations of total transparency through the barrier. Latter, in 2015 [40] the same authors extended the previous study to the quantum tunneling of electrons in bilayer graphene. In this case, they showed that the statistical complexity can take its minimum value even when the transmission through the barrier is near zero, which is an evidence that the statistical complexity is capable to distinguish certain physical differences between similar physical systems.

From Eqs. (21), (26) and (28), the Shannon entropy and the disequilibrium read, respectively

$$S_q = n\gamma + n + \frac{1}{2} + \ln(\sqrt{\hbar\pi}\, n!\, 2^n \rho) - 2\sum_{k=1}^{n} {}_2F_2\left(1,1;\frac{3}{2},2;-x_{n,k}^2\right) x_{n,k}^2$$

$$+ \sum_{k=1}^{n}\sum_{i=1}^{n} \binom{n}{i}\frac{(-1)^i\, 2^i}{i}\, {}_1F_1\left(1;\frac{1}{2};-x_{n,k}^2\right), \tag{29}$$

$$D_q = \frac{1}{\rho\sqrt{\hbar}} \sum_{j=0}^{2n} \frac{\Gamma\left(j+\frac{1}{2}\right)}{2^{j+\frac{1}{2}}} \frac{4!}{(2j+4)!} B_{2j+4,4}\left(c_0^{(n)}, 2!\, c_1^{(n)}, \ldots, (2j+1)!\, c_{2j}^{(n)}\right), \tag{30}$$

where $\gamma$ is the Euler constant, $x_{n,k}$ ($k = 1, 2, \ldots, n$) are the roots of $H_n(x)$, ${}_1F_1$ and ${}_2F_2$ are the hypergeometric functions [41], and

$$c_l^{(n)} = \begin{cases} \dfrac{(-1)^{\frac{3n-l}{2}} n! \, 2^{l-1}}{l! \left(\frac{n-l}{2}\right)! \sqrt{2^n \, n! \sqrt{\pi}}} [(-1)^l + (-1)^n], & 0 \leq l \leq n \\ 0, & l > n \end{cases} \quad (31)$$

$$B_{m,l}(a_1, a_2, \ldots, a_{m-l+1}) = \sum_{\hat{\pi}(m,l)} \frac{m!}{j_1! j_2! \cdots j_{m-l+1}!} \left(\frac{a_1}{1!}\right)^{j_1} \left(\frac{a_2}{2!}\right)^{j_2} \cdots$$

$$\times \left(\frac{a_{m-l+1}}{(m-l+1)!}\right)^{j_{m-l+1}}, \quad (32)$$

where the sum runs over all the partitions $\hat{\pi}(m,l)$ such that $j_1 + j_2 + \cdots + j_{m-l+1} = l$ and $j_1 + 2j_2 + \cdots + (m-l+1)j_{m-l+1} = m$. Surprisingly, we can see from Eqs. (27), (29) and (30) that the statistical complexity remain time-independent for any state $n$ and does not depend on the conductivity $\sigma(t)$ of the superconductor.

## 3. Results and discussion

As we can see from Eqs. (22)-(25) and (29)-(32), all the statistical measures are written in terms of $\rho(t)$, the solutions of the generalized Milne-Pinney. By considering the conductivity $\sigma(t) = \sigma_0/(At + 1)$ and following the procedure decribed in Refs. [42-44], we obtain

$$\rho(t) = \sqrt{\frac{\pi}{2A}} (At + 1)^{1/2\left(1 - \frac{\sigma_0}{A\varepsilon_0}\right)} \left[J_\beta^2(k(At + 1)) + Y_\beta^2(k(At + 1))\right]^{1/2}, \quad (33)$$

where $J_\beta(x)$ and $Y_\beta(x)$ are the first- and second-kind Bessel functions, respectively, and

$$\beta = \frac{1}{2}\sqrt{1 + \frac{2\sigma_0}{A\varepsilon_0} + \frac{\sigma_0^2}{A^2\varepsilon_0^2}}, \quad (34)$$

$$k = \frac{c}{\lambda_L A}. \quad (35)$$

The Figs. 1(a)-(b) show that $D_q(t)$ ($H_q(t)$) increases (decreases) more rapidly with increasing time for greater conductivity values $\sigma_0$. This result can be explained as follows. The charge inside the superconductor decreases, as can be seen in Fig. 2 and Fig. 3, the probability density becomes more localized for greater conductivity $\sigma_0$ (Fig. 2), and with increasing time (Fig. 3). The results show that $H_q(t)$ and $D_q(t)$ can also be used to determine the moment when all the charge that was in the superconductor was expelled.

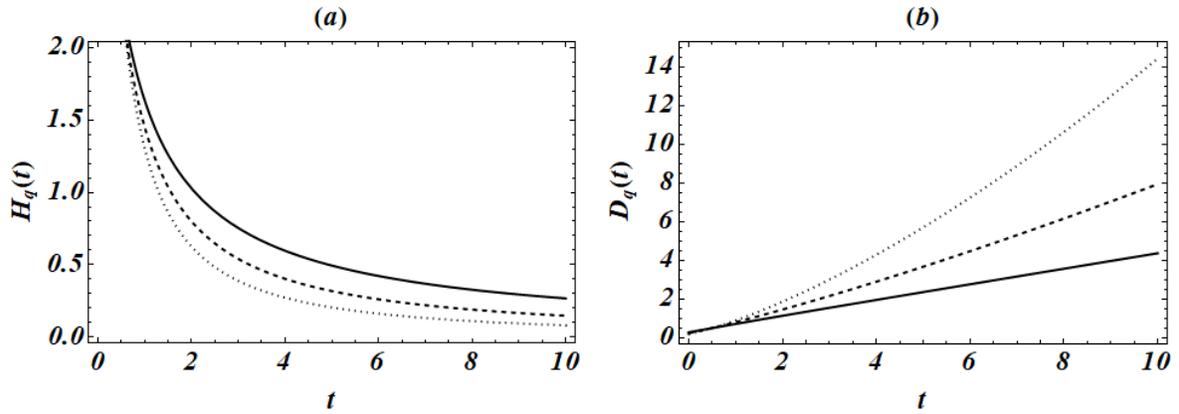

**Figure 1.** Plots of (a) $H_q(t)$ and (b) $D_q(t)$ for $\sigma_0 = 2$ (solid line), $\sigma_0 = 2.5$ (dashed line) and $\sigma_0 = 3$ (dotted line). In this figure we consider $A = \varepsilon_0 = c = \lambda_L = \hbar = 1$.

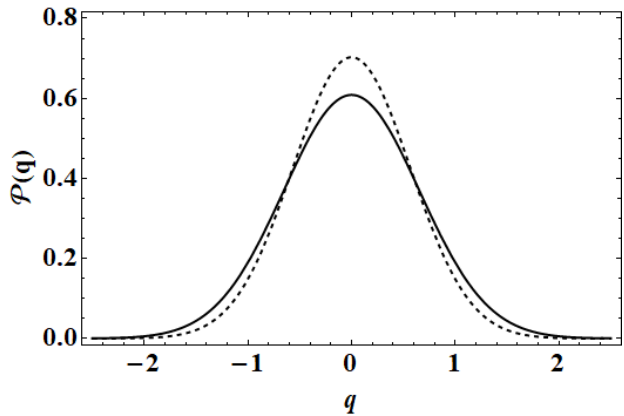

**Figure 2.** Plots of $\mathcal{P}(q)$ of the ground state for $\sigma_0 = 0.5$ (solid line) and $\sigma_0 = 3$ (dashed line). In this figure we consider $A = \varepsilon_0 = c = \lambda_L = \hbar = 1$ and $t = 0.5$.

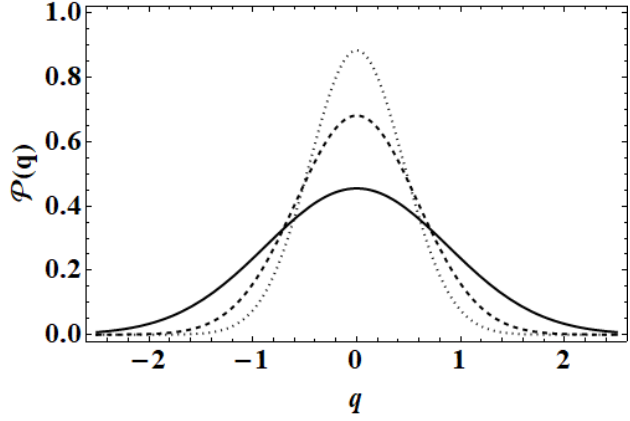

**Figure 3.** Plots of Plots of $\mathcal{P}(q)$ of the ground state for $t = 0$ (solid line) and $t = 0.5$ (dashed line) and $t = 1$ (dotted line). In this figure we consider $A = \varepsilon_0 = c = \lambda_L = \hbar = 1$ and $\sigma_0 = 1.5$.

For the system described by the time-dependent Schrödinger equation (10), the expectation value of the energy armazened in the superconductor is given by

$$\langle E_n \rangle = \left[ \frac{1}{2L^2(t)\rho^2(t)} \left(1 + L^2(t)\rho^2(t)\dot{\rho}^2(t)\right) + \frac{\omega^2(t)\rho^2(t)}{2} \right] \left(n + \frac{1}{2}\right), \quad (36)$$

using Eq. (33), we show in Fig. 4 that as the charge is expelled the energy goes rapidly to zero for any quantum state.

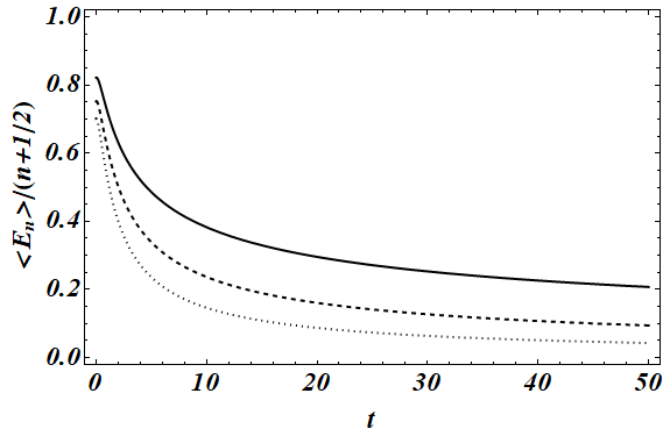

**Figure 4.** Plots of $\langle E_n \rangle / (n + 1/2)$ for $\sigma_0 = 0.4$ (solid line), $\sigma_0 = 0.6$ (dashed line) and $\sigma_0 = 0.8$ (dotted line). In this figure we consider $A = \varepsilon_0 = c = \lambda_L = \hbar = 1$.

## 4. Concluding remarks

In this work we introduced an alternative quantization scheme for the charge within a superconductor in terms of two canonically conjugated operator, the charge operator, $q$, and the magnetic flux operator, $\Phi$. Following the London approach, we show that a time-dependent Hamiltonian (Eq.(9)) leads to the classical equation of motion for the charge within of a superconductor. This allowed us to introduce the quantum Hamiltonian in terms of the operators $q$ and $\Phi$. Using the Lewis and Riesenfeld invariant method, we obtain the solutions of the Schrödinger equation in the charge space of a superconductor in the two-fluid model with time-dependent conductivity.

From the wave-functions obtained, we calculated the time-dependent uncertainties and the mean energy. Information measures were also obtained, such as Shannon entropy and complexity. The later is always time-independent and also does not depend on conductivity. The others quantities are written in terms of a time-dependent function, $\rho(t)$, the solution of Milne-Pinney equation (13). These are interesting results because to find the wave-functions, uncertainties, Shannon entropy, the mean energy and the desequilibrium of any time-dependent superconductor system described by Eq. (10), one has only to solve the respective Milne–Pinney equation.

**Acknowledgments**

The authors are grateful to the Coordenação de Aperfeiçoamento de Pessoal de Nível Superior (CAPES) and the National Counsel of Scientific and Technological Development (CNPq) of Brazil for financial support.